%% file: paper_itise2021_review_time_series_python.tex
\documentclass[runningheads]{llncs}

\usepackage[utf8]{inputenc} 

\usepackage{geometry} 
\usepackage{graphicx} 
\usepackage{tikz}
\usetikzlibrary{shapes.geometric, arrows}
\usepackage{longtable}
\usepackage{pdflscape}
\usepackage{array} 
\usepackage{url}
\usepackage[hidelinks]{hyperref}
\usepackage{subcaption}
\usepackage{multirow}

\title{A systematic review of \python{} packages for time series analysis}
\author{Julien Siebert\inst{1}\orcidID{0000-0002-7696-0046} \and
Janek Groß\inst{1}\orcidID{0000-0002-6306-711X} \and Christof Schroth\inst{1}\orcidID{0000-0003-0828-4063}}

\institute{Fraunhofer Institut for Experimental Software Engineering IESE.\\Fraunhofer Platz 1, 67663 Kaiserslautern, Germany}

\newcommand{\pypi}{\href{https://pypi.org}{PyPI}}
\newcommand{\pip}{\href{https://pip.pypa.io/en/stable/}{pip}}
\newcommand{\conda}{\href{https://docs.conda.io/en/latest/}{conda}}
\newcommand{\github}{\href{https://github.com}{GitHub}}
\newcommand{\python}{Python}
\begin{document}
\maketitle
\begin{abstract}
This paper presents a systematic review of \python{} packages with a focus on time series analysis. The objective is to provide (1) an overview of the different time series analysis tasks and preprocessing methods implemented, and (2) an overview of the development characteristics of the packages (e.g., documentation, dependencies, and community size). This review is based on a search of literature databases as well as \github{} repositories. Following the filtering process, 40 packages were analyzed. We classified the packages according to the analysis tasks implemented, the methods related to data preparation, and the means for evaluating the results produced (methods and access to evaluation data). We also reviewed documentation aspects, the licenses, the size of the packages' community, and the dependencies used. Among other things, our results show that forecasting is by far the most frequently implemented task, that half of the packages provide access to real datasets or allow generating synthetic data, and that many packages depend on a few libraries (the most used ones being numpy, scipy and pandas). We hope that this review can help practitioners and researchers navigate the space of \python{} packages dedicated to time series analysis. We will provide an updated list of the reviewed packages online at \url{https://siebert-julien.github.io/time-series-analysis-python/}.
 
\end{abstract}

\section{Introduction}
A time series is a set of data points generated from successive measurements over time. The analysis of this type of data has found application in many fields, from finance to health, including the monitoring of computer networks or the environment. The current trend of reducing the cost of sensors and data storage, the increasing performance of Big Data and data analysis technologies such as machine learning or data mining, are opening up more and more possibilities to acquire and analyze temporal data. 
Moreover, as the number of time series analysis application cases rises, more and more data scientists, data engineers, analysts, and software engineers have to use dedicated time series analysis libraries.

In this article, we systematically review \python{} packages dedicated to time series analysis. \python{} is one of the programming languages of choice for data scientists\footnote{See the survey performed by Kaggle in 2018: \url{https://www.kaggle.com/kaggle/kaggle-survey-2018}.}. Data scientists are not only responsible for analyzing data; their task is also to ensure that services based on these analyses reach a sufficient level of maturity to be deployed and maintained in production. In this context, we review not only the analysis tasks implemented in the packages, but also several factors external to the tasks themselves, such as which dependencies are used or how big the community behind the development of the package in question is. Our goal is not to evaluate the quality of the implementations themselves but to provide a structured overview that is useful for data scientists confronted with time series analysis (and faced with having to choose which packages to rely on), the scientific community, and the community of Python developers working in this field. This paper is structured as follows: Related work is introduced in section \ref{sect.rel-work}; the search methodology and the search results are described in sections \ref{sect.methodo} and \ref{sect.results}, respectively; threats to validity are discussed in section \ref{sect.validity}; and section \ref{sect.conclusion} concludes the paper.

\section{Related Work}\label{sect.rel-work}
Time series analysis is a broad research field covering many application domains. The literature contains many reviews, either focusing on analysis tasks and methods (see, for instance, these reviews on forecasting \cite{Hendikawati.2020,Mahalakshmi.2016,Panigrahi.2020,Tealab.2018}, clustering and classification \cite{Abanda.2019,Aghabozorgi.2015,Bagnall.2017,Fawaz.2019,Susto.2018}, anomaly detection \cite{Ayadi.2017,Cook.2020,Wu.2016}, changepoint analysis \cite{Aminikhanghahi.2017,Sharma.2016,Truong.2020}, pattern recognition \cite{Torkamani.2017,Wang.2018}, or dimensionality reduction \cite{S.Badhiye.2018}) or focusing on a specific application domain (see, for instance, these surveys on finance \cite{Sezer.2020}, IoT and Industry 4.0 \cite{Lepenioti.2020,Mohammadi.09.12.2017,Zhao.2019}, or health \cite{Zeger.2006}). Over time, several formal definitions and reviews of time series analysis tasks have been published; see, for example \cite{Esling.2012,Fakhrazari.2017,Keogh.2003}.

However, existing implementations (software packages or libraries) are often listed - usually in a non-systematic way - in textbooks (like \cite{Cowpertwait.2009,Shumway.2017} for R, or \cite{Nielsen.2019} for \python{}) or gray literature (for example, Towards Data Science\footnote{\url{https://towardsdatascience.com/}}, KDnuggets\footnote{\url{https://www.kdnuggets.com/}} or Machine Learning Mastery\footnote{\url{https://machinelearningmastery.com/}}), and few papers actually systematically review packages or libraries in a specific language. For example, \cite{Joo.2020} reviewed packages for analyzing animal movement data in R, and \cite{Slater.2019} surveyed R packages for hydrology. With respect to \python{}, we found several reviews of packages for different domains: social media content scrapping \cite{Thivaharan.2020}, topological data analysis \cite{Ray.2018}, or data mining \cite{Stancin.2019}. For time series analysis in \python{}, the only related work we could find is \cite{Januschowski.2019}, where the authors review packages focusing on forecasting.

There is, to the best of our knowledge, no systematic review of \python{} packages for generic time series analysis.

\section{Methodology}\label{sect.methodo}
\begin{figure}%
\centering
\input{images/flow.tex}
\caption{Search and filtering process overview. Edge labels indicate the number of repositories left after each step.}%
\label{fig.flow}%
\end{figure}
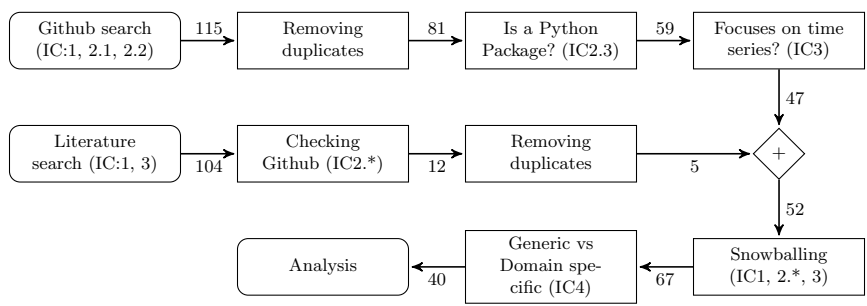

We conducted a systematic literature review according to \cite{Kitchenham.2013}. However, these guidelines focus on printed literature, not on software packages. Hence, we adjusted these methods.
Our search process is illustrated in Figure \ref{fig.flow}. We conducted a search in both literature databases and code repositories (\github{}). The following sections provide more details on the different steps of the search itself.

\subsection{Research questions}
We already stated our goal and the context we set for this review in the introduction. We formalize this context as follows: We want to \textbf{analyze} \python{} packages dedicated to time series analysis \textbf{for the purpose of} structuring the available implementations (we explicitly exclude the purpose of evaluating them) \textbf{with respect to} the implemented time series analysis tasks \textbf{from the viewpoint of} practitioners \textbf{in the context of} building data-driven services on top of these implementations. Our research questions are:

\begin{itemize}
	\item \textbf{RQ1} Which time series analysis tasks exist? And which of these are implemented in maintained Python packages?
	\item \textbf{RQ2} How do the packages support the evaluation of the produced results? 
	\item \textbf{RQ3} How do the packages support their usage, and what insights can we gain to estimate the durability of a given package and make an informed choice about its long-term use?
\end{itemize}

\subsection{Inclusion criteria}
To guide our review and filter relevant packages, we defined the following inclusion criteria (IC):
The package should be open source, written in \python{}, and available on \github{} (\textbf{IC1}). The package should be actively maintained (last commit within less than 6 months) (\textbf{IC2.1}); it should have more than 100 \github{} stars (\textbf{IC2.2}); and it should be listed in \pypi{}\footnote{PyPI is the \python{} Package Index, a repository of software for the \python{} programming language, see \url{https://pypi.org/}.} and be installable via \pip{}\footnote{pip is the \python{} Package Installer, see \url{https://pip.pypa.io/en/stable/}} or \conda{}\footnote{conda is a \python{} package management system and environment management provided by the Anaconda distribution, see \url{https://docs.conda.io/en/latest/}.} (\textbf{IC2.3}). The package should explicitly target time series analysis (\textbf{IC3}). We excluded packages that can be used for time series analysis (as building blocks) but whose main purpose is not time series analysis per se (for example, generic scientific computing packages such as \href{https://www.scipy.org/}{scipy} or \href{https://numpy.org/}{numpy}, packages dedicated to data manipulation or storage such as \href{https://pandas.pydata.org/}{pandas}, or generic machine learning or data mining packages such as \href{https://scikit-learn.org}{scikit-learn}). Finally, we focused our search on packages offering methods that tend to be domain-agnostic (\textbf{IC4}) and excluded domain-specific packages.
Domain-specific packages are packages aiming to solve time series analysis in a specific domain (for example, audio, finance, geoscience, etc.). They usually focus on specific types and formats of time series and domain related analysis tasks. 

\subsection{Searching open-source repositories in \github{}}
In order to filter \github{} repositories, we selected a list of topics\footnote{\url{https://github.com/topics}}, filtered the results by language (\python{}, IC1), by number of stars (at least 100, IC2.2), and considered only repositories that were updated after July 2020 (IC2.1).

In order to select a list of relevant topics, we first manually selected a list of eight \python{} packages known to be used in time series analysis (i.e., a seeds set): pandas, numpy, scipy, statsmodel, ruptures, tsfresh, tslearn, and sktime; as well as a sample of the packages using the topic "time-series".
We examined the topics used by these packages and then extended this list of topics with different spellings while manually double-checking their existence in \github{}. We considered a total of 16 different topics (see Table \ref{tab.topics}). 
The first search led to a total of 115 repositories.

\begin{table}%
	\centering
	\begin{tabular}{|c|c|c|c|}
		\hline
		\href{https://github.com/topics/time-series?l=python}{time-series} & \href{https://github.com/topics/time-series-regression?l=python}{time-series-regression} & \href{https://github.com/topics/signal-processing?l=python}{signal-processing} & \href{https://github.com/topics/time-series-classification?l=python}{time-series-classification}\\
		
		\href{https://github.com/topics/time-series-analysis?l=python}{time-series-analysis} & \href{https://github.com/topics/time-series-forecast?l=python}{time-series-forecast} & \href{https://github.com/topics/time-series-visualization?l=python}{time-series-visualization} & \href{https://github.com/topics/time-series-decomposition?l=python}{time-series-decomposition}\\
		
		\href{https://github.com/topics/time-series-forecasting?l=python}{time-series-forecasting} & \href{https://github.com/topics/time-series-data-mining?l=python}{time-series-data-mining} & \href{https://github.com/topics/timeseries?l=python}{timeseries} & \href{https://github.com/topics/timeseries-forecasting?l=python}{timeseries-forecasting}\\
		
		\href{https://github.com/topics/time-series-prediction?l=python}{time-series-prediction} & \href{https://github.com/topics/time-series-segmentation?l=python}{time-series-segmentation} & \href{https://github.com/topics/timeseries-analysis?l=python}{timeseries-analysis} & \href{https://github.com/topics/time-series-clustering?l=python}{time-series-clustering}\\
		\hline
	\end{tabular}
\caption{List of topics used to conduct the search on \github{}}
\label{tab.topics}
\end{table}

\subsubsection{Removing duplicates}
We found 24 unique repositories that were duplicated (i.e., listed in more than one topic). 
After duplicate removal, 81 unique repositories remained. 

\subsubsection{Checking if the repository contains the code of a \python{} package}
We restricted our search to packages that are referenced by \pypi{} and can be installed with \pip{} or \conda{} (IC2.3).
Note that the repository name might not reflect the package name (if one exists). For example, the repository \url{https://github.com/PyWavelets/pywt} contains the source code for the package named pywavelets. The repository \url{https://github.com/angus924/rocket} does not contain the source code for the \python{} package rocket. We therefore checked each of the 81 repositories manually and excluded 22 repositories, which yielded a total of 59 remaining repositories that contain the source code of a \python{} package.

\subsubsection{Including only packages focused on time series analysis}
Finally, we manually checked whether the focus of the package is time series analysis (IC3).
After exclusion, 47 remaining packages were kept for further analysis.

\subsection{Searching scientific bibliographic databases}
The search for packages only in a repository might not be sufficient to cover all existing packages. For example, one of our seed packages (namely tsfresh) was not uncovered by the search. Hence, we extended our search to existing literature and software databases. 
We used the bibliographic databases IEEE Xplore\footnote{\url{https://ieeexplore.ieee.org}}, ACM Digital Library\footnote{\url{https://dl.acm.org/}}, Web of Science\footnote{\url{https://www.webofknowledge.com}}, and Scopus \footnote{\url{https://www.scopus.com/}}, as well as the Journal of Open Source Software (JOSS)\footnote{\url{https://joss.theoj.org/}}, and Zenodo\footnote{\url{https://zenodo.org/}}.
For IEEE Xplore, ACM Digital Library, Web of Science, and Scopus, we limited ourselves to the search string \verb|"Python" AND "time series"| in the document title. For the Journal of Open Source Software (JOSS), we first used the key words \verb|"time series"| and then filtered the results by language (the query used is: \url{https://joss.theoj.org/papers/search?q=time+series}). For Zenodo, we also used the search string \verb|"Python" AND "time series"|, limited the search to the software category and removed the duplicates (e.g., different versions of the same software). The full query for Zenodo is: 
\url{https://zenodo.org/search?page=1&size=200&q=%22time%20series%22%20AND%20%22python%22&sort=mostrecent&type=software}.
We only included references that matched our inclusion criteria IC1, IC2.*, and IC3. Table \ref{table-literature-search} summarizes our search results.

\begin{table}
	\centering
		\begin{tabular}{|c|c|c|c|}
			\hline
			Data Source & Number of hits & Number of included documents & Included references\\
			\hline
			IEEE Xplore&1&0 &\\
			ACM Digital Library&2&1&\cite{DavidBurns.2018}\\
			Web of Science&10&4&\cite{Christ.2018,Alexandrov.2020,DavidBurns.2018,Faouzi.2020} \\
			Scopus&12&4&\cite{Christ.2018,Alexandrov.2020,DavidBurns.2018,Faouzi.2020}\\
			JOSS&21&1&\cite{Law.2019}\\
			Zenodo&68&6&\cite{Collenteur.2020,Miller.2019,Scholzel.2019,SilvaPetronioCandidoDeLimaE.2019,Snow.2020,Team.2016}\\
			\hline
		\end{tabular}
	\caption{Literature search results}
	\label{table-literature-search}
\end{table}

We manually cross-checked the results obtained from \github{} with the results obtained by our literature search. Out of the eleven packages resulting from our literature search, only five repositories were not already in the \github{} search results: \href{https://github.com/blue-yonder/tsfresh}{tsfresh}, \href{https://github.com/neurodsp-tools/neurodsp}{neurodsp}, \href{https://github.com/repos/springer-math/Mathematics-of-Epidemics-on-Networks}{EoN}, \href{https://github.com/repos/CSchoel/nolds}{nolds}, and \href{https://github.com/repos/pastas/pastas'}{pastas}.

\subsection{Snowballing}
In order to extend our search, we used a snowballing approach. We first manually reviewed the package documentations in order to find links to other similar packages. Only two packages --- tsfresh\footnote{\url{https://tsfresh.readthedocs.io/en/latest/text/introduction.html\#what-else-is-out-there}} and sktime\footnote{\url{https://www.sktime.org/en/latest/related_software.html}} --- actually document related packages. Second, we manually reviewed the documentation and the \github{} repositories of all packages to find related publications. We then reviewed the papers to find new packages (i.e., we performed a single backward snowballing pass). Out of a total of 79 packages, 15 new packages were included after the snowballing phase, for a total of 67 packages.

\subsection{Generic vs. domain-specific packages (\textbf{IC4})}
Finally, we classified the packages in two categories: domain-specific and generic. As previously defined, we consider domain-specific packages to be packages aiming to solve time series analysis in a specific domain (for example, audio, finance, geoscience, etc.) and generic packages as those offering methods that tend to be domain-agnostic. Out of the 67 packages, 27 packages were categorized as domain-specific and 40 packages as generic.

\subsection{Data extraction and categorization}
We manually extracted relevant information about the packages from their documentation pages and code. For the categorization, we used an iterative, bottom-up approach. Two researchers first proposed category definitions and then categorized the packages. A third researcher was responsible for resolving disagreements. Iterations were performed until the categories and results were consolidated. 

\section{Results}\label{sect.results}
\subsection{RQ1: Implementation of the time series analysis tasks}
To answer our research question RQ1, we first reviewed the task definitions present in the literature and then analyzed the 40 packages classified as generic to extract information about which tasks have been implemented in the packages.

\subsubsection{Task definitions}
Time series analysis tasks are formally defined in the literature. Reviews like \cite{Esling.2012,Fakhrazari.2017,Fu.2011,Keogh.2003} define the following tasks: \textbf{Indexing (query by content)}: given a time series and some similarity measure, find the nearest matching time series \cite{Esling.2012,Fakhrazari.2017,Keogh.2003}.
\textbf{Clustering}: find groups (clusters) of similar time series \cite{Esling.2012,Fakhrazari.2017,Fu.2011,Keogh.2003}.
\textbf{Classification}: assign a time series to a predefined class \cite{Esling.2012,Fakhrazari.2017,Fu.2011,Keogh.2003}.
\textbf{Segmentation (Summarization)}: create an accurate approximation of a time series by reducing its dimensionality while retaining its essential features \cite{Esling.2012,Fakhrazari.2017,Fu.2011,Keogh.2003}.
\textbf{Forecasting (Prediction)}: given a time series dataset up to a given time $t_n$, forecast the next values \cite{Esling.2012,Fakhrazari.2017}.
\textbf{Anomaly Detection}: find abnormal data points or subsequences (also called discords) \cite{Esling.2012,Fakhrazari.2017}.
\textbf{Motif Discovery}: find every subsequence (called motif) that appears recurrently in a time series \cite{Esling.2012,Fakhrazari.2017,Fu.2011}.
\textbf{Rules Discovery (Rule Mining)}: find the rules that may govern associations between sets of time series or subsequences \cite{Fakhrazari.2017,Fu.2011}.

Esling and Agon also define implementation components \cite{Esling.2012}: \textbf{preprocessing} (e.g., filtering noise, removing outliers, or imputing missing values), \textbf{representation} (e.g., dimensionality reduction, finding fundamental shape characteristics), \textbf{similarity measures}, and \textbf{indexing schemes}.

\subsubsection{Implemented tasks}
While analyzing the packages, we found packages explicitly mentioning the tasks corresponding to our literature review. We found 20 packages explicitly providing forecasting methods (T1), 6 packages providing classification methods (T2), 6 packages providing clustering methods (T3), 6 packages providing anomaly detection methods (T4), and 4 packages providing segmentation methods (T5). We classified four packages under the category \textit{pattern recognition} (T6), encompassing both indexing and motif discovery tasks. We also classified five packages under the category \textit{change point detection} (T7), which was not in our literature review. Finally, we could not find any package explicitly mentioning the rules discovery task.

Considering the implementation components, we found 4 packages explicitly providing \textit{dimensionality reduction} methods (DP1), 17 packages explicitly providing \textit{missing values imputation} methods (DP2), 16 packages explicitly providing \textit{decomposition} methods (e.g., decomposing time series into trends, seasonal components, or frequency components) (DP3), 24 packages explicitly providing generic \textit{transformation and features generation} methods (DP4), and 7 packages explicitly providing methods for computing similarity measures (DP5). Table \ref{tab.classif} gives an overview of our categorization of the packages.

Forecasting is by far the most frequently implemented task. There is no significant difference, in terms of number of packages, between the other tasks. However, we need to be cautious when interpreting these numbers. First, the tasks as formally defined in the literature might not be explicitly mentioned in the packages documentation or code. Second, the delineation between a task and the methods used to implement it is sometimes blurry and context dependent. For example, one can perform change point detection for the sake of finding time points where some time series properties change and, as a consequence, raising alarms in a production system, or use it as a preprocessing step for segmenting a time series into different phases. Another example are forecasting models, which can also be applied for outlier detection.

\subsection{RQ2: Evaluation of the produced results}
To answer our research question RQ2, we extracted information about the evaluation of the outcomes produced by the packages. We came up with two main clusters: functions that facilitate the evaluation itself (E1, E2, E3) and functions for either generating synthetic data or downloading existing datasets (D1, D2). We found 13 packages explicitly providing methods for model selection, hyperparameter search, or feature selection (E1), 20 packages explicitly providing evaluation metrics and statistical tests (E2), and 25 packages providing visualization methods (E3). Concerning the data, we found 16 packages explicitly providing functions for generating synthetic time series data (D1), and 19 packages providing access to time series datasets (D2). A large majority of the packages provide a way to evaluate the results produced. Only 4 packages have not been classified in any of the E or D classes.

\subsection{RQ3: Package usage and community}
\begin{figure}
\begin{subfigure}[t]{0.49\textwidth}
	\centering
	\includegraphics[height=5cm]{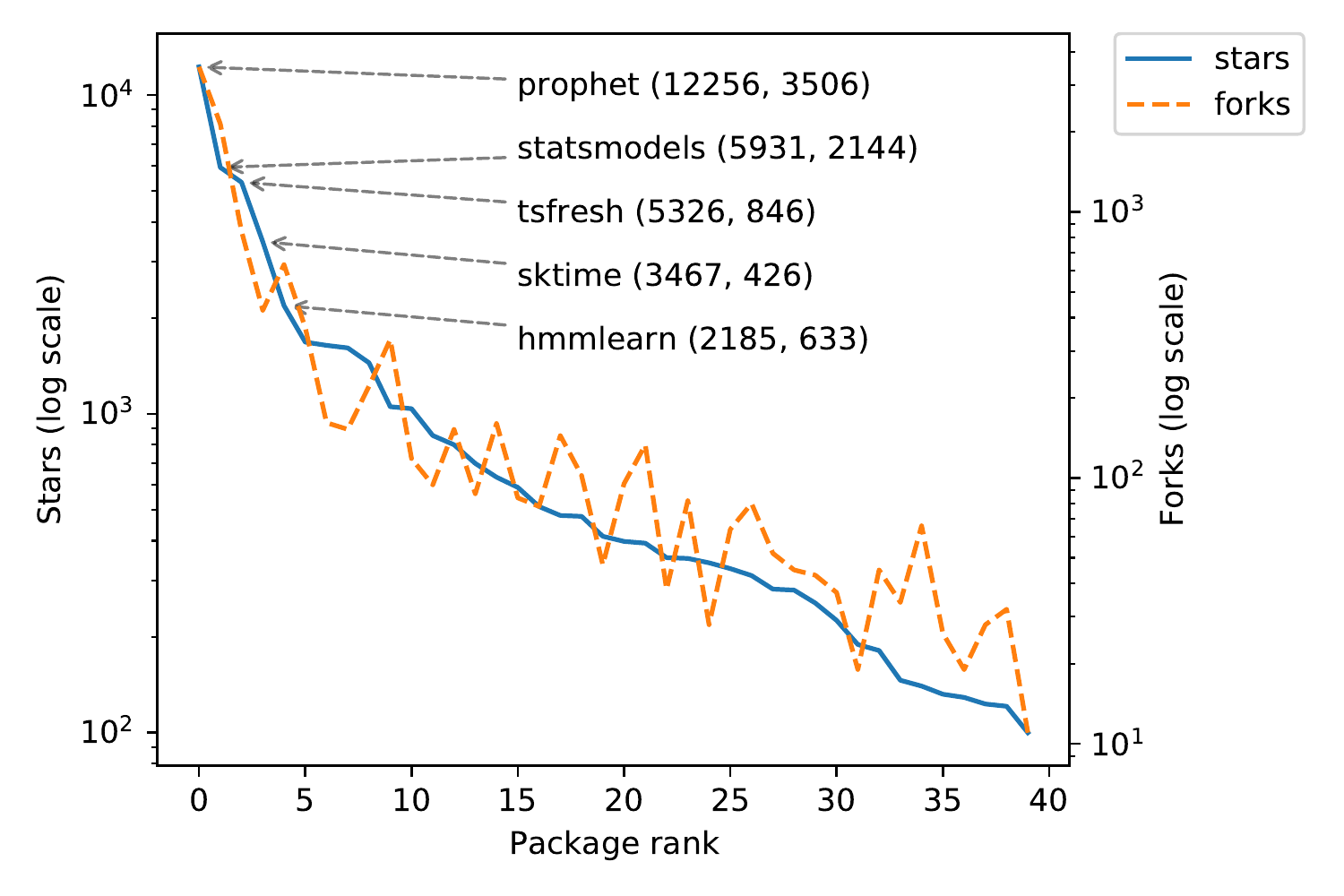}
	\caption{Distribution of stars and forks for all 40 repositories (log scale). The repositories are ranked by the number of stars, in descending order.\label{fig.nb-stars}}
\end{subfigure}
\hspace{1pt}
\begin{subfigure}[t]{0.49\textwidth}
	\centering
	\includegraphics[height=5cm]{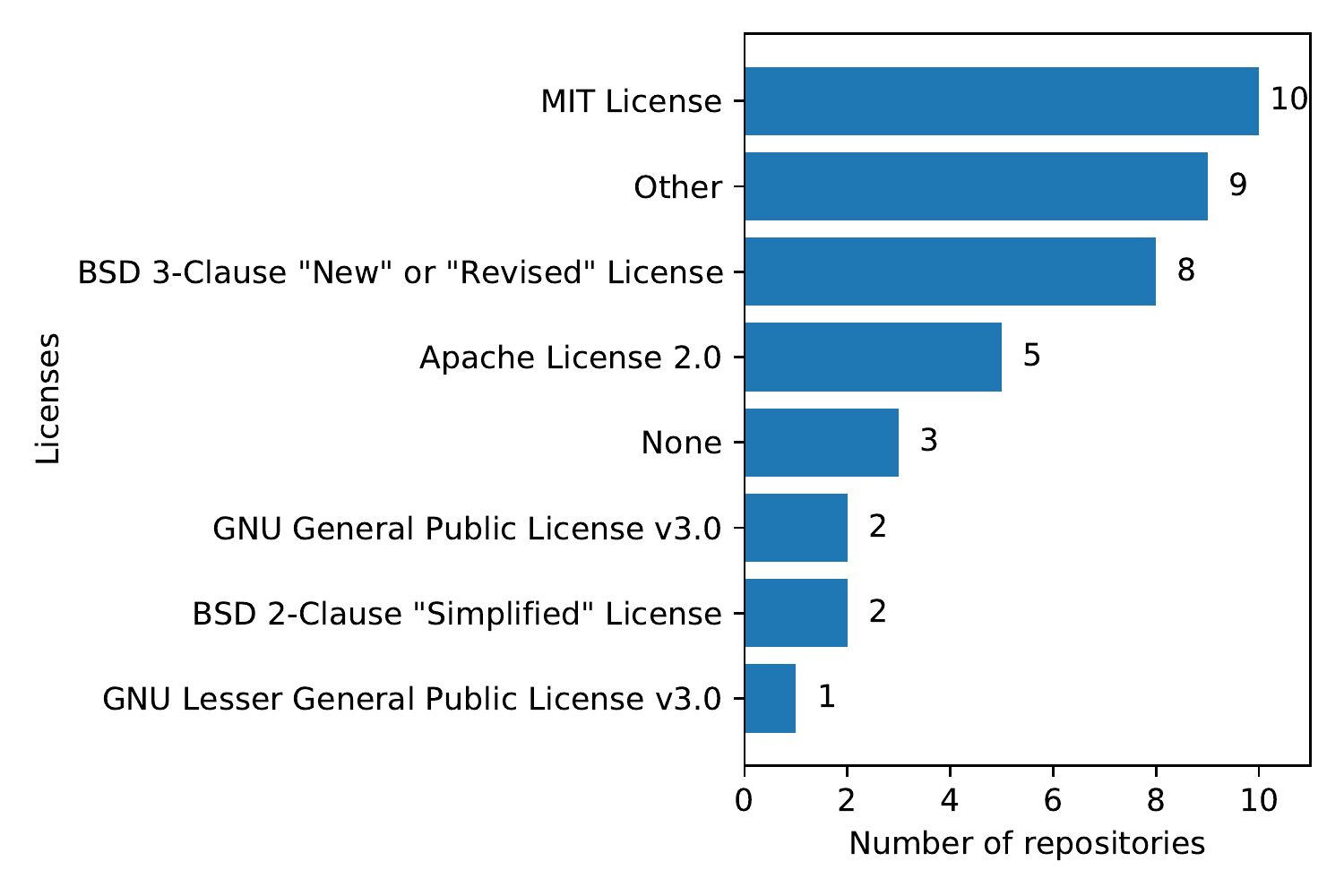}
	\caption{Distribution of licenses (number of repositories per license). None means that no license information was available from \github{} directly. \label{fig.nb-licenses}}
\end{subfigure}
\end{figure}

To answer our research question RQ3, we extracted information about the documentation, the dependencies, and the community supporting the packages. For instance, \github{} provides many statistics about a repository (e.g., the number of stars, forks, issues) that can be used to get a first idea of the liveliness of the different packages. We used the number of \github{} stars and forks to estimate the community behind each package. Figure \ref{fig.nb-stars} shows the distribution of stars and forks for all 40 packages. Another piece of information that is relevant to practitioners are the licenses under which the implementations are available. Figure \ref{fig.nb-licenses} shows the distribution of the licenses used among the 40 repositories.

We also investigated the dependencies used by each of the selected 40 packages. We used the \python{} program johnnydep\footnote{\url{https://pypi.org/project/johnnydep/}} to automatically collect the dependencies without installing the packages directly. We only looked at direct dependencies required for the installation of the package. We did not consider specific installation options such as dev or test. We did not search for all dependencies recursively.
Here is an example of how we called the program \verb|johnnydep| \verb|PACKAGENAME| \verb|--fields=ALL| \verb|--no-deps| \verb|--output-format=json|. The dependencies of two packages could not be retrieved automatically (cesium and deeptime). We also manually cross-checked the dependencies and filled in the missing ones. Table \ref{table-most-used-dependencies} shows which dependencies are used the most by the packages.

Almost all packages (37) depends upon numpy. The top 5 dependencies are \href{https://numpy.org/}{numpy}, \href{https://www.scipy.org/}{scipy} (scientific computing), \href{https://pandas.pydata.org/}{pandas} (data manipulation), \href{https://scikit-learn.org}{scikit-learn} (machine learning), and \href{https://matplotlib.org}{matplotlib} (visualization).

\begin{table}
	\centering
		\begin{tabular}{|c|c|c||c|c|c|}
			\hline
			Package Name & Used & Rank & Package Name & Used & Rank \\
			\hline
numpy       & 37 & 1 & torch       & 6 & 8 \\
scipy       & 30 & 2 & numba       & 6 & 8 \\
pandas       & 23 & 3 & cython       & 6 & 8 \\
scikit-learn   & 21 & 4 & tensorflow    & 5 & 9 \\
matplotlib    & 16 & 5 & seaborn      & 4 & 10 \\
statsmodels   & 8 & 6 & future       & 4 & 10 \\
tqdm        & 7 & 7 & joblib       & 4 & 10 \\
			\hline
		\end{tabular}
	\caption{Ranking of the most frequently used dependencies}
	\label{table-most-used-dependencies}
\end{table}

Finally, we investigated five documentation aspects (Do1-Do5). We found that 30 packages provide a separate documentation page (Do1). The other ten packages use the README of the repository file as documentation. 18 packages provide notebooks directly executable without installation via a link to either mybinder.org\footnote{\url{https://mybinder.org/}} or Google Colab\footnote{\url{https://colab.research.google.com/}} (Do2 +), 12 packages provide stand-alone notebook files to be downloaded (Do2 *), and 10 packages do not provide any notebook file at all. 28 packages provide an API reference (Do3). All packages provide an installation page (Do4) and almost all packages (38) provide user guides in the form of static examples or tutorials.

\begin{table}%
	\resizebox{\textwidth}{!}{%
	\begin{tabular}{|c|c|c|c|c|c|c|c||c|c|c|c|c||c|c|c||c|c||c|c|c|c|c|}
		\hline
		\multirow{2}{4em}{\textbf{Package Name}} & \multicolumn{7}{c||}{\textbf{Tasks}} & \multicolumn{5}{c||}{\textbf{Data Preparation}} & \multicolumn{3}{c||}{\textbf{Evaluation}} & \multicolumn{2}{c||}{\textbf{Data}} & \multicolumn{5}{c|}{\textbf{Documentation}} \\
		\cline{2-23}
										&T1 &T2 &T3 &T4 &T5 &T6 &T7 &DP1&DP2&DP3&DP4&DP5&E1 &E2 &E3 &D1 &D2 &Do1&Do2&Do3&Do4&Do5\\
		\hline
		arch						& + &   &   &   &   &   &   &   &   &   &   &   &   & + &   & + & + & + & * & + & + & +\\
		\hline
		atspy						& + &   &   &   &   &   &   &   & + & + & + &   & + & + & + &   &   &   & + &   & + & +\\
		\hline
		banpei					&   &   &   & + &   &   & + &   &   &   &   &   &   &   &   &   &   &   &   &   & + & +\\
		\hline
		cesium					&   &   &   &   &   &   &   &   &   &   & + &   &   &   &   &   &   & + & * & + & + & +\\
		\hline
		darts						& + &   &   &   &   &   &   &   & + & + & + &   &   & + & + & + & + & + &   & + & + & +\\
		\hline
		deeptime				& + &   & + &   &   &   &   & + &   & + & + &   &   & + & + & + &   & + & + &   & + & +\\
		\hline
		deltapy					& + &   & + &   &   &   &   & + &   & + & + & + &   &   &   &   &   & + & + & + & + & +\\
		\hline
		dtaidistance		&   &   & + &   &   &   &   &   &   &   &   & + &   &   & + &   &   & + & + & + & + & +\\
		\hline
		EMD-signal			&   &   &   &   &   &   &   &   &   & + &   &   &   &   & + &   &   & + & + & + & + & +\\
		\hline
		flood-forecast	& + &   &   &   &   &   &   &   & + &   & + &   &   & + & + &   &   & + & + & + & + & +\\
		\hline
		gluonts					& + &   &   &   &   &   &   &   & + &   & + &   &   & + & + & + & + & + &   & + & + & +\\
		\hline
		hcrystalball		& + &   &   &   &   &   &   &   & + &   & + &   & + & + & + & + & + & + & * & + & + & +\\
		\hline
		hmmlearn				& + &   &   &   &   &   &   &   &   &   & + &   &   &   &   & + &   & + & * &   & + & +\\
		\hline
		hypertools			&   &   & + &   &   &   &   & + & + &   & + &   &   &   & + &   &   & + & * & + & + & +\\
		\hline
		linearmodels		&   &   &   &   &   &   &   &   &   &   &   &   &   & + &   &   & + &   & * &   & + & +\\
		\hline
		luminaire				& + &   &   & + &   &   & + &   & + & + & + &   & + &   &   &   &   & + &   &   & + & \\
		\hline
		matrixprofile		&   &   & + & + & + & + &   &   & + &   &   & + &   &   & + &   & + & + &   & + & + & +\\
		\hline
		mcfly						&   & + &   &   &   &   &   &   &   &   &   &   & + &   & + &   &   & + &   &   & + & +\\
		\hline
		neuralprophet		& + &   &   &   &   &   &   &   & + & + & + &   &   &   & + &   &   & + & * & + & + & +\\
		\hline
		nolds						&   &   &   &   &   &   &   &   & + &   &   &   &   & + &   & + & + & + &   & + & + & +\\
		\hline
		pmdarima				& + &   &   &   &   &   &   &   &   & + & + &   & + & + & + &   & + &   & * &   & + & +\\
		\hline
		prophet					& + &   &   &   &   &   & + &   &   & + &   &   &   & + & + &   &   & + & * & + & + & +\\
		\hline
		pyaf						& + &   &   &   &   &   &   &   & + & + & + &   & + & + & + &   & + & + &   & + & + & +\\
		\hline
		pycwt						&   &   &   &   &   &   &   &   &   & + &   &   &   &   &   &   & + & + & + & + & + & +\\
		\hline
		pydlm						& + &   &   &   &   &   &   &   &   & + &   &   & + & + & + &   &   & + & * & + & + & +\\
		\hline
		pyFTS						& + &   &   &   &   &   &   &   &   &   & + &   & + & + & + & + & + &   & + &   & + & +\\
		\hline
		pyodds					&   &   &   & + &   &   &   &   &   &   &   &   & + & + & + &   &   & + & * & + & + & +\\
		\hline
		pytorchts				& + &   &   &   &   &   &   &   & + &   & + &   &   &   & + & + & + & + & + & + & + & +\\
		\hline
		pyts						&   & + &   &   & + & + &   &   & + & + & + & + &   & + & + & + & + & + & * & + & + & \\
		\hline
		PyWavelets			&   &   &   &   &   &   &   &   &   & + & + &   &   &   &   & + & + & + & * & + & + & +\\
		\hline
		ruptures				&   &   &   &   &   &   & + &   &   &   &   &   &   & + & + & + &   & + & + &   & + & +\\
		\hline
		scikit-multiflow&   & + &   & + &   &   & + &   & + &   & + &   &   & + & + & + &   & + & * & + & + & \\
		\hline
		seglearn				&   &   &   &   &   &   &   &   &   &   & + &   &   &   &   &   & + & + & * & + & + & +\\
		\hline
		sktime					& + & + &   & + & + &   &   &   & + & + & + & + & + & + & + &   & + & + & * & + & + & +\\
		\hline
		sktime-dl				& + & + &   &   &   &   &   &   &   &   &   &   & + &   &   &   &   &   & + &   & + & +\\
		\hline
		statsmodels			& + &   &   &   &   &   &   & + & + & + &   &   & + & + & + & + & + & + & * & + & + & +\\
		\hline
		stumpy					&   &   &   &   & + & + &   &   &   &   &   & + &   &   &   &   &   & + & * & + & + & +\\
		\hline
		tftb						&   &   &   &   &   &   &   &   &   & + & + &   &   & + & + & + &   & + & + & + & + & +\\
		\hline
		tsfresh					&   &   &   &   &   &   &   &   & + &   & + &   & + & + &   &   & + & + &   & + & + & +\\
		\hline
		tslearn					&   & + & + &   &   & + &   &   &   &   & + & + &   & + &   & + & + & + &   &   & + & +\\
		\hline
		\textbf{Total}& \textbf{20} & \textbf{6} & \textbf{6} & \textbf{6} & \textbf{4} & \textbf{4} & \textbf{5} & \textbf{4} & \textbf{17} & \textbf{16} & \textbf{24} & \textbf{7} & \textbf{13} & \textbf{23} & \textbf{25} & \textbf{16} & \textbf{19} & \textbf{34} & \textbf{30} & \textbf{28} & \textbf{40} & \textbf{37}\\
		\hline
		\multirow{2}{4em}{} & T1 & T2 & T3 & T4 & T5 & T6 & T7 & DP1 & DP2 & DP3 & DP4 & DP5 & E1 & E2 & E3 & D1 & D2 & Do1 & Do2 & Do3 & Do4 & Do5 \\
		\cline{2-23}
		 & \multicolumn{7}{c||}{\textbf{Tasks}} & \multicolumn{5}{c||}{\textbf{Data Preparation}} & \multicolumn{3}{c||}{\textbf{Evaluation}} & \multicolumn{2}{c||}{\textbf{Data}} & \multicolumn{5}{c|}{\textbf{Documentation}} \\
		\hline
	\end{tabular}
	} 
	\caption{Classification of packages. Tasks: T1 (forecasting), T2 (classification), T3 (clustering), T4 (anomaly detection), T5 (segmentation), T6 (pattern recognition), T7 (change point detection). Data Preparation (also called implementation components): DP1 (dimensionality reduction), DP2 (missing values imputation), DP3 (decomposition), DP4 (preprocessing), DP5 (similarity measures). Evaluation: E1 (model selection, hyperparameter search, feature selection), E2 (metrics and statistical tests), E3 (visualization). Datasets: D1 (synthetic data generation) and D2 (contains datasets). Documentation: Do1 (dedicated documentation), Do2 (notebook: directly executable (+), present (*)), Do3 (API reference), Do4 (install guide), Do5 (user guide).}
	\label{tab.classif}
\end{table}

\section{Discussion and threats to validity}\label{sect.validity}

In this section, we discuss the choices we made and that may affect the validity of this review.

This review focused on \github{}. Gitlab and Sourceforge were checked manually, but we decided not to include them as sources due to the insufficient number of results. 

We limited ourselves to packages with at least 100 stars. This somehow arbitrary limit led us to exclude packages with a number of stars close to 100 (e.g., the \href{https://github.com/StingraySoftware/stingray}{stingray} package with 93 stars at the time of the search). We excluded packages that were not maintained but might have been relevant for practitioners. An example is the \href{https://github.com/RJT1990/pyflux}{pyflux} package (forecasting). We also excluded repositories that are not \python{} packages. This led us to discard interesting repositories like \href{https://github.com/shubhomoydas/ad_examples}{ad\_examples} (which provides state-of-the-art anomaly detection methods) and many repositories containing code scripts associated with scientific papers. 

Concerning the search process, we used a mix of literature databases and \github{} topics together with a snowballing approach to find relevant packages. The reason forthis was that several known packages could not be found automatically. For example, the package \href{https://github.com/cesium-ml/cesium}{cesium} does not list any topic and therefore was not found in our first \github{} search. It was found after snowballing. Another example is \href{https://github.com/blue-yonder/tsfresh}{tsfresh}, which was missing in the first \github{} search and was found in the literature search. The problem may be the language filter (strictly \python{}), as \href{https://github.com/blue-yonder/tsfresh}{tsfresh} lists some of the topics we searched for ("time-series").

We tried to automate some of the tasks (e.g., filtering repositories that contain \python{} packages or finding the dependencies), using both \pypi{} and \github{} API, or the johnnydep tool. There were false positives and false negatives. This led us to manually cross check the results obtained from our automated search.

Whether a package focuses on time series analysis or not can sometimes be fuzzy. For example, we decided to leave the topic of survival analysis out of this review. We initially found two packages: \href{https://github.com/CamDavidsonPilon/lifelines}{lifelines} and \href{https://github.com/sebp/scikit-survival}{scikit-survival}. The same applies to the boundary between generic and domain-specific packages. We took a conservative approach to keep our survey sufficiently focused.

As already mentioned above, the definition of what should be regarded as a task vs. an "implementation component" is difficult, as a strict boundary may not even exist. Moreover, it is sometimes not clear what methods the packages provide without actually installing them and testing them. Indeed, the documentation might not be complete or the vocabulary used may differ from one package to another. One solution was to check the code itself. Here again, the search strings used play an important role in avoiding false negatives.

\section{Conclusion}\label{sect.conclusion}
This paper presented a systematic review of \python{} packages dedicated to time series analysis. The search process led to a total of 40 packages that were analyzed further. We proposed a categorization of the packages based on the analysis tasks implemented, the methods related to data preparation, the means for evaluating the results produced, and the kind of documentation present, and also looked at some development aspects (licenses, stars, dependencies). We also discussed the search process with its possible bias and the challenges we encountered while searching for and reviewing the relevant packages. The scope of this survey does, however, not include any evaluation of the implementations or the results they would produce, for example, on benchmark datasets. We hope that this review can help practitioners and researchers navigate the space of \python{} packages dedicated to time series analysis. Since the packages will evolve, we plan to maintain an updated list of the reviewed packages online at \url{https://siebert-julien.github.io/time-series-analysis-python/}.

\section*{Acknowledgments} The authors want to thank Anna Maria Vollmer and Markus L\"{o}ning for their valuable feedback. 

\bibliographystyle{splncs04}
\bibliography{bibliography}

\end{document}

%% file: images/flow.tex
\tikzstyle{startstop} = [rectangle, rounded corners, minimum width=3cm, minimum height=1cm, text centered, draw=black, text width=2.5cm]
\tikzstyle{process} = [rectangle, minimum width=3cm, minimum height=1cm, text centered, draw=black, text width=2.5cm]
\tikzstyle{decision} = [diamond, minimum width=0.5cm, minimum height=0.5cm, text centered, draw=black]

\begin{tikzpicture}[
    pre/.style={=stealth',semithick},
    post/.style={->,shorten >=1pt,>=stealth',semithick},
		scale=0.75, every node/.style={scale=0.75}
    ]
\node (git-search) [startstop] at (0,2) {Github search (IC:1, 2.1, 2.2)};
\node (git-duplicate) [process] at (4,2) {Removing duplicates};
\node (git-package) [process] at (8,2) {Is a Python Package? (IC2.3)};
\node (git-focus) [process] at (12,2) {Focuses on time series? (IC3)};

\node (lit-search) [startstop] at (0,0) {Literature search (IC:1, 3)};
\node (lit-filter) [process] at (4,0) {Checking Github (IC2.*)};
\node (lit-duplicate) [process] at (8,0) {Removing duplicates};

\node (merge) [decision] at (12,0) {$+$};

\node (snowballing) [process] at (12,-2) {Snowballing (IC1, 2.*, 3)};
\node (filter-gen-dom) [process] at (8,-2) {Generic vs Domain specific (IC4)};
\node (analysis) [startstop] at (4,-2) {Analysis};

\draw[post] (git-search)--(git-duplicate) node [midway, above] {115}; 
\draw[post] (git-duplicate)--(git-package) node [midway, above] {81}; 
\draw[post] (git-package)--(git-focus) node [midway, above] {59}; 

\draw[post] (lit-search)--(lit-filter) node [midway, below] {104};
\draw[post] (lit-filter)--(lit-duplicate) node [midway, below] {12};

\draw[post] (git-focus)--(merge) node [midway, right] {47};
\draw[post] (lit-duplicate)--(merge) node [midway, below] {5};

\draw[post] (merge)--(snowballing) node [midway, right] {52};
\draw[post] (snowballing)--(filter-gen-dom) node [midway, below] {67};
\draw[post] (filter-gen-dom)--(analysis) node [midway, below] {40};

\end{tikzpicture}